\documentclass[sigconf,printcopyright=false]{acmart}
\usepackage{booktabs} % For formal tables
\newcommand{\xhdr}[1]{\vspace{1.7mm}\noindent{{\bf #1.}}}

\copyrightyear{}
\acmYear{2019}
\setcopyright{none}

\acmConference[WWW '19]{Proceedings of the 2019 World Wide Web Conference}{May 13--17, 2019}{San Francisco, CA, USA}
\acmBooktitle{Proceedings of the 2019 World Wide Web Conference (WWW'19), May 13--17, 2019, San Francisco, CA, USA}
\acmPrice{}
\acmDOI{10.1145/3308558.3313512}
\acmISBN{978-1-4503-6674-8/19/05}
\usepackage{balance}

\begin{document}
\title{Predicting pregnancy using large-scale data\\from a women's health tracking mobile application}

\author{Bo Liu$^{\dagger}$}
\affiliation{%
  \institution{Dept. of Computer Science, Stanford}
}

\author{Shuyang Shi$^{\dagger}$}
\affiliation{%
  \institution{Dept. of Computer Science, Stanford}
}
\author{Yongshang Wu$^{\dagger}$}
\affiliation{%
  \institution{Dept. of Computer Science, Stanford}
}

\author{Daniel Thomas}
\affiliation{%
  \institution{Clue by BioWink GmbH, Berlin}
}

\author{Laura Symul}
\affiliation{%
  \institution{Dept. of General Surgery and Dept. of Statistics, Stanford}
}

\author{Emma Pierson$^*$}
\affiliation{%
  \institution{Dept. of Computer Science, Stanford}
}

\author{Jure Leskovec}
\affiliation{%
  \institution{Dept. of Computer Science, Stanford}
    \institution{Chan-Zuckerberg Biohub}
}

\renewcommand{\shortauthors}{B. Liu et al.}

\begin{abstract}
Predicting pregnancy has been a fundamental problem in women's health for more than 50 years. Previous datasets have been collected via carefully curated medical studies, but the recent growth of women's health tracking mobile apps offers potential for reaching a much broader population. However, the feasibility of predicting pregnancy from mobile health tracking data is unclear. Here we develop four models -- a logistic regression model, and 3 LSTM models -- to predict a woman's probability of becoming pregnant using data from a women's health tracking app, Clue by BioWink GmbH. Evaluating our models on a dataset of 79 million logs from 65,276 women with ground truth pregnancy test data, we show that our predicted pregnancy probabilities meaningfully stratify women: women in the top 10\% of predicted probabilities have a 89\% chance of becoming pregnant over 6 menstrual cycles, as compared to a 27\% chance for women in the bottom 10\%. 
We develop a technique for extracting \emph{interpretable} time trends from our deep learning models, and show these trends are consistent with previous fertility research.
Our findings illustrate the potential that women's health tracking data offers for predicting pregnancy on a broader population; we conclude by discussing the steps needed to fulfill this potential.

\end{abstract}

\keywords{Pregnancy prediction; mobile health tracking}

\maketitle

\section{Introduction}

Predicting pregnancy is a fundamental problem in women's health~\citep{dunson2001bayesian,dunson2005bayesian,lum2016bayesian,scarpa2007bayesian}. Modeling the probability of pregnancy, and identifying behaviors which affect it, allows for more efficient family planning. Identifying people with a very low chance of becoming pregnant is important for targeting and treating infertility, which can be a devastating experience~\citep{greil1997infertility}. Because of the importance of predicting and facilitating pregnancy, the fertility services market is predicted to grow to \$21 billion by 2020~\citep{fertility_services_market_fast_company}. 

Attempts to predict pregnancy go back decades, often using Bayesian methods ~\citep{dunson2001bayesian,dunson2005bayesian,lum2016bayesian,scarpa2007bayesian} to capture the complex dynamics which govern fertility. Many factors influence fertility, including age~\citep{dunson2004infertility} and medical conditions~\citep{kamel2010infertilecouple,moos2008healthier}. Fertility fluctuates over the course of the menstrual cycle (which averages 28-29 days in length, although this varies across women~\citep{chiazze1968length}), peaking during the ``fertile window'', around 14 days before the end of the cycle. For women with a regular cycle of 28-29 days, the fertile window occurs approximately midway through the cycle~\citep{wilcox2000timing,barrett1969risk}. Because the timing of the fertile window varies, and intercourse during the fertile window is much more likely to result in pregnancy, fertility prediction methods often focus on detecting and predicting the fertile window, using ovulation tests and measurements of cervical mucus and basal body temperature~\citep{fehring2002accuracy,bigelow2004mucus}. 

Previous studies of fertility have relied on carefully curated data from medical studies~\citep{dunson2001bayesian,dunson2005bayesian,lum2016bayesian,scarpa2007bayesian,barrett1969risk,schwartz1980fecundability,royston1982basal,royston1999new}: for example, participants are generally asked to regularly monitor critical features like basal body temperature and menstrual cycle starts and to take pregnancy tests after each menstrual cycle. Many people are unwilling or unable to engage in such careful monitoring, limiting the applicability of these methods to the general population. 

In the last few years, the growing use of health tracking mobile apps offers a new potential data source for predicting pregnancy in broader populations. Health tracking apps have been recognized as delivering a ``data bounty'' \citep{hayden2016mobile} for healthcare~\citep{wilbanks2016stop, o2017mobile}. Within women's health specifically, health tracking apps are used by millions of women worldwide \citep{clue_nyt,epstein2017examining, Chakradhar2018}, and have already been used to study sexually transmitted infections ~\citep{alvergne2018sexually}, menstrual cycle fluctuations~\citep{pierson2018modeling,symul2018assessment}, and menstrual cycle lengths~\citep{hillard2017data}. These apps allow users to create profiles which include age and birth control information and keep daily logs of symptoms relevant to pregnancy, including basal body temperature, period starts, and ovulation tests. Crucially, they also allow women to log positive and negative pregnancy tests, providing a ground truth source of pregnancy data. Pregnancy data collected by women's health mobile apps differs from that collected in medical studies in three ways: 

\begin{enumerate} 
\item \textbf{Larger datasets.} The dataset we consider in this paper contains tens of millions of observations from tens of thousands of women, orders of magnitude more than previous datasets~\citep{ecochard2006heterogeneity}. Complex models developed for these smaller datasets, which often rely on computationally intensive techniques like MCMC sampling, will not scale to datasets from mobile health tracking apps, necessitating new methods.
\item \textbf{Broader populations.} While previous fertility studies have predominantly focused on filtered, mostly healthy or proven fertile populations that volunteer for research within a single country, health tracking apps can be used by anyone with a smartphone, and are used by women all over the world. 
\item \textbf{Missing data.} Critical features like basal body temperature (BBT) and pregnancy tests are less reliably recorded in mobile health tracking datasets than in medical studies, making missing data a pressing concern. 
\end{enumerate}

Mobile tracking datasets thus present new challenges but also new opportunities for predicting pregnancy in a much broader population. To this end, a number of companies have announced initiatives to predict pregnancy or fertility using mobile tracking data~\citep{glowprediction,naturalcyclesprediction,scherwitzl2017perfect,kindaraprediction}. To date, however, it is unclear whether their fertility prediction algorithms are reliable, because they are proprietary (so they cannot be compared to or independently assessed) and their efficacy has been disputed~\citep{naturalcycles1,naturalcycles2,naturalcycles3}. 

\xhdr{This work} Here we present what is to our knowledge the first study of pregnancy prediction using fully described methods applied to large-scale women's health tracking app data. Using a dataset from a women's health tracking app which includes ground truth data on pregnancy tests, we develop and assess four pregnancy prediction models --- an interpretable logistic regression model, and three LSTM models --- which integrate traditional fertility modeling methods and modern time series prediction methods. We first show that our models meaningfully stratify women by their probability of becoming pregnant: women in the top 10\% of predicted probabilities have a 89\% chance of becoming pregnant over 6 menstrual cycles, while women in the bottom 10\% have only a 27\% chance. 
We further provide an intuitive technique for extracting \emph{interpretable} time trends from our LSTM models, overcoming a common failing of deep learning models, and show that these time trends are consistent with previous fertility research. 
Finally, we discuss the steps that should be taken to maximize the efficacy of pregnancy prediction from women's health tracking data. 

\vspace{-3mm}
\section{Problem setup} 

Our task is predicting, on the basis of the health tracking data a woman logs during a single menstrual cycle, whether she will log a positive pregnancy test at the end of that cycle. We first describe our dataset and then describe our task in more detail. 

\subsection{Dataset} We use data from the Clue women's health tracking app~\citep{clue_data_citation}. The app has been rated the most accurate menstrual cycle tracking app by gynecologists~\citep{moglia2016evaluation} and previously used in studies which show that it reliably replicates known women's health findings~\citep{alvergne2018sexually,hillard2017data,pierson2018modeling}. All data is de-identified and analysis was determined to be exempt from review by the Stanford Institutional Review Board.

\xhdr{Features} Our dataset consists of two types of features: 
\begin{itemize} 
\item \textbf{User features.} Users can record age and birth control method (e.g., ``None'', ``Condoms'', ``IUD'', or various types of birth control pill). We encode age using a two-element vector: a binary element which indicates whether age data is missing, and a continuous element which is the value of age if it is present, and 0 otherwise. We encode birth control using a vector with indicator variables for each type of birth control; the vector is all zeros if birth control data is missing. 
\item \textbf{Daily feature logs.} Each row in this dataset consists of one log of one feature for one user on one date. Users can log 110 \emph{binary} features, including positive/negative pregnancy tests (which we use to define ground truth, and do not include in the inputs to our model), period bleeding, mood and behavior features (eg, happy mood or exercise:running), and taking daily birth control. Users can also log 3 \emph{continuous} features: basal body temperature (BBT), resting heart rate, and body weight. We list all daily features in Table \ref{all_features}.
\end{itemize} 

\begin{table*}[h]
\small
  \caption{All daily features which appear in our data. Each feature has both a category and a type. Features are binary except for those in the ``continuous'' category. While pregnancy tests are included in the binary features, we do not use them as predictive features. HBC=hormonal birth control; TNP=type not provided.}
  \label{all_features}
  \begin{tabular}{p{2.5cm}p{10cm}}
\toprule
              \textbf{Category} &                                           \textbf{Type} \\
\midrule
            Ailment &                                        Allergy, Cold/Flu Ailment, Fever, Injury \\
       Appointment &                                                  Date, Doctor, Ob Gyn, Vacation \\
 Collection Method &                                         Menstrual Cup, Pad, Panty Liner, Tampon \\
        Continuous &                                                 BBT, Resting Heart Rate, Weight \\
           Craving &                                                  Carbs, Chocolate, Salty, Sweet \\
         Digestion &                                      Bloated, Gassy, Great Digestion, Nauseated \\
           Emotion &                                                      Happy, PMS, Sad, Sensitive \\
            Energy &                                   Energized, Exhausted, High Energy, Low Energy \\
          Exercise &                                                 Biking, Running, Swimming, Yoga \\
             Fluid &                                             Atypical, Creamy, Egg White, Sticky \\
              Hair &                                                            Bad, Dry, Good, Oily \\
     Injection HBC &                                                 Administered, Type Not Provided \\
               IUD &                            Inserted, Removed, Thread Checked, TNP \\
        Medication &                            Antibiotic, Antihistamine, Cold/Flu Medication, Pain \\
            Mental &                                             Calm, Distracted, Focused, Stressed \\
        Motivation &                                Motivated, Productive, Unmotivated, Unproductive \\
              Pain &                                Cramps, Headache, Ovulation Pain, Tender Breasts \\
             Party &                             Big Night Party, Cigarettes, Drinks Party, Hangover \\
         Patch HBC &               Removed, Removed Late, Replaced, Replaced Late, TNP\\
            Period &                                                  Heavy, Light, Medium, Spotting \\
          Pill HBC &                                  Double, Late, Missed, Taken, TNP \\
              Poop &                                            Constipated, Diarrhea, Great, Normal \\
          Ring HBC &               Removed, Removed Late, Replaced, Replaced Late, TNP \\
               Sex &                              High Sex Drive, Protected, Unprotected, Withdrawal \\
              Skin &                                                           Acne, Dry, Good, Oily \\
             Sleep &                             0-3 Hrs, 3-6 Hrs, 6-9 Hrs, 9 Hrs, TNP \\
            Social &                                       Conflict, Sociable, Supportive, Withdrawn \\
              Test &  Ovulation Neg, Ovulation Pos, Pregnancy Neg, Pregnancy Pos \\
\bottomrule
\end{tabular}
\end{table*}

\xhdr{Data filtering} To ensure that users are regularly using the app, we filter for users who enter at least 300 daily feature logs. We apply basic quality control filters to the three continuous features to remove unreliable values (for example, BBT $>110^\circ F$). To minimize the chance that users are already pregnant, we filter out all cycles after a user has logged a positive pregnancy test. 

\xhdr{Defining menstrual cycle starts} Because fertility fluctuates over the course of the menstrual cycle, defining menstrual cycle starts -- when the period begins -- is necessary to predict pregnancy. 
We define a \emph{cycle start} as a start of bleeding after the user has not recorded any bleeding for at least 7 days. So, for example, if the user records bleeding on May 1, 2, and 29, the cycle starts would be May 1 and May 29. A user's \emph{cycle day} is the number of days since their most recent cycle start, with 0 denoting the day of cycle start. 

\xhdr{Encoding continuous features}
\label{continuous_features_section}
We encode continuous features (eg, weight) using a two-element vector: a binary element which indicates whether data is present or missing, and a continuous element which is the value for the feature if present, and 0 otherwise. For each cycle and each user, we subtract off the mean for each continuous feature (since the \emph{change} in features like BBT is most indicative of cycle phase and therefore fertility~\citep{buxton1948hormonal}). We include each user's average value of the three continuous features as a user-specific feature. 

\begin{figure}[h!]
    \centering
    \includegraphics[width=0.47\textwidth]{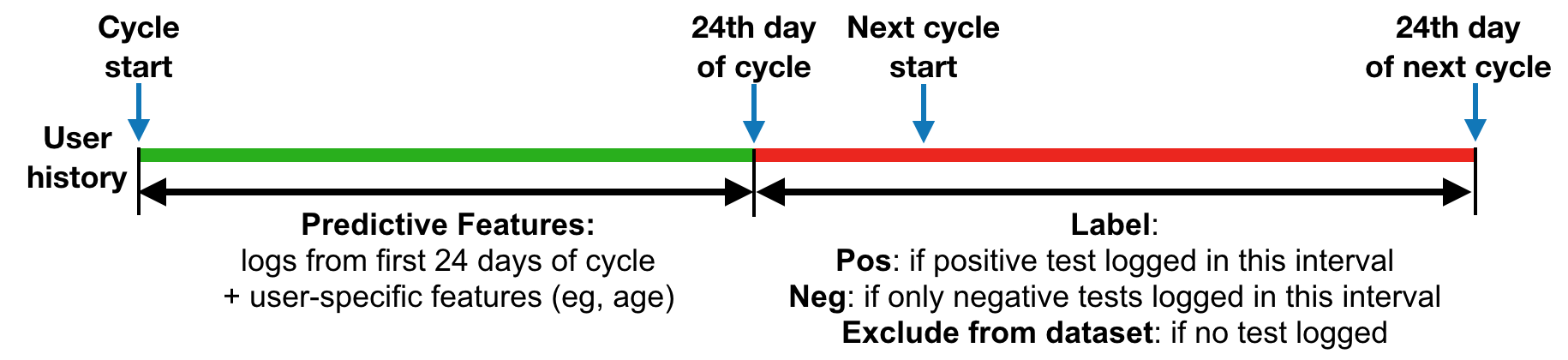}
    \vspace{-2mm}
    \caption{Prediction task. The model makes predictions using logs from the first 24 days of a cycle (green interval), and the cycle is labeled using pregnancy tests taken after day 24 of the cycle and before day 24 of the next cycle (red interval). The vast majority of pregnancy tests in our dataset are taken near when the user's cycle is supposed to start, consistent with proper use of pregnancy tests, so any positive pregnancy tests likely result from activity during the green interval, which will be included in the feature vector.}
    \label{fig:prediction-task}
\end{figure}

\subsection{Prediction task} Our task is predicting from the first 24 days of a user's cycle whether she will become pregnant in that cycle. We choose the 24-day interval because most women are very unlikely to become pregnant past this interval~\citep{wilcox2000timing}. Our prediction task is binary, and each example is one cycle for one woman. We define positive examples as cycles in which the woman logs a positive pregnancy test after day 24 of the cycle and before day 24 of the next cycle (Figure~\ref{fig:prediction-task}); we define negative examples as cycles with a negative pregnancy test and no positive pregnancy test. Thus, our dataset consists only of cycles followed by positive or negative pregnancy tests\footnote{We define positive and negative examples in this way to mitigate missing data concerns. The alternative (i.e. assuming that any cycle without a positive pregnancy test is a negative example) is problematic for two reasons. First, some users may not bother to log positive tests, so the negative label is a false negative. Second, some users may not bother to log birth control, concealing the fact that they have very little chance of getting pregnant---consistent with this, we find that the fraction of positive examples using this method, even for users who log no birth control, is far lower than the previous literature would imply.}. Importantly, predicting whether a test will be positive or negative is a difficult and useful prediction task, because it occurs in settings where the woman herself is sufficiently uncertain about whether she is pregnant that she believes taking a test is worthwhile. Our dataset consists of 16,580 positive cycles, 88,685 negative cycles, 65,276 women, and 79,423,281 symptom logs. The proportion of cycles which are positive in our dataset (16\%) is consistent with previous work~\citep{colombo2000daily}. 
We use data from 90\% of users for training/validation and report results on a test set of the remaining 10\% of users. We train on a dataset balanced for positive and negative examples by downsampling each training batch (a standard technique for training on unbalanced data~\citep{japkowicz2000class}). Following standard practice, we report all results on the unbalanced test dataset. 

\section{Previous work}

Our predictive models rely on ideas both from the fertility modeling literature and from the deep learning time series prediction literature. We now summarize work in both areas. 

\subsection{Fertility modeling}

\label{fertility_modeling_lit_review}
We briefly summarize three main approaches in the extensive fertility modeling literature; ~\citep{ecochard2006heterogeneity} provides a lengthier review.

\begin{enumerate}

\item \textbf{Time-to-pregnancy (TTP) models}~\citep{tietze1987differential, potter1960length, sheps1964time,heckman1990testing} assume each individual menstrual cycle results in pregnancy with some probability $\mu$ which can vary across couples and over time. These models are useful for capturing the high level covariates which influence $\mu$ (eg, age) but they do not capture detailed daily dynamics within a single cycle (eg, sex during the fertile window is more likely to result in pregnancy). Therefore, they are less useful for modeling the health-tracking datasets we consider, whose strength is their detailed daily information.

\item \textbf{Barrett-Marshall and Schwartz (BMS)  models}~\citep{barrett1969risk, schwartz1980fecundability,royston1982basal,dunson2001bayesian,dunson2005bayesian,stanford2007effects} allow for more detailed modeling of daily activity within each cycle. They assume each act of sex contributes independently to the probability a cycle results in pregnancy. 
The probability that someone becomes pregnant in a cycle is thus
\begin{flalign*}
    &1 - \prod_d (1-f_d)^{S_d},
\end{flalign*}
where $d$ is the cycle day, $f_d$ is the probability of getting pregnant by having sex only on day $d$, and $S_d$ is a binary variable indicating whether sex occurred on day $d$. The term inside the product is the probability that day $d$ does \emph{not} result in pregnancy; it is 1 if $S_d$ is 0, indicating that no sex occurred, and $1 - f_d$ if $S_d$ is 1. Schwartz ~\citep{schwartz1980fecundability} extends this model by assuming that the overall probability of pregnancy also depends on a couple-specific parameter which captures, for example, the fact that some couples are infertile irrespective of sexual activity patterns. However, this additional parameter creates model identifiability problems. 

\item \textbf{Extended Time-to-Pregnancy (ETTP) models}~\citep{clayton1997artificial,royston1999new} provide an approximation to the BMS model by assuming that only sex on the most fertile day contributes to the probability of pregnancy, and sex on other days has no effect. As this model is only a pragmatic approximation to the true generative process, we instead develop a computationally efficient implementation of the BMS model. 
\end{enumerate}

All three of the above models were developed for datasets orders of magnitude smaller than the one we consider. Consequently, they perform parameter inference by estimating the posterior distribution, often using computationally intensive techniques like MCMC, and will not scale to our dataset. Therefore, in this work, we extend the BMS model using a scalable deep-learning-based model. Our extension scales to health-tracking datasets because it relies on more efficient backpropagation-based optimization. 

\subsection{Deep learning for time series prediction}  

Long Short Term Memory (LSTM) neural networks~\citep{hochreiter1997long} have been successfully used to model medical time series data~\citep{baytas2017patient, lipton2015learning, pham2016deepcare}. They maintain a hidden state at each timestep which can be used for prediction, and are a natural choice for prediction on time series with discrete timesteps (in our case, days of the menstrual cycle). Because they rely on more scalable backpropagation methods, they also scale to large health-tracking datasets.

\section{Models}

We develop four models. We select model hyperparameters -- learning rate, hidden size, number of layers, batch size, dropout rate, and regularization strength -- using grid search. Our models are publicly available at \url{ https://github.com/AndyYSWoo/pregnancy-prediction}. 

\begin{figure}[h!]
    \centering
    \includegraphics[width=0.23\textwidth]{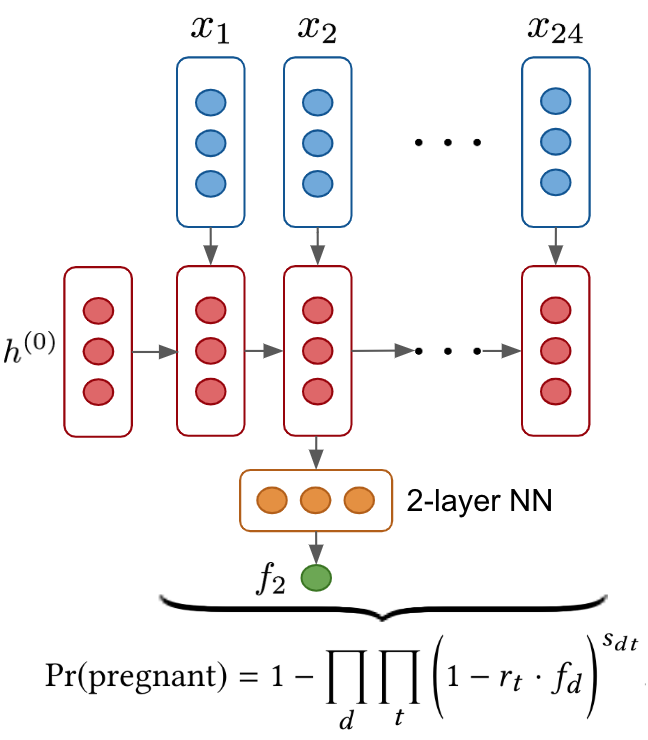}
    \caption{LSTM + BMS fertility model: we feed the daily features $x$ (blue) into an LSTM model to obtain the hidden state $h$ (red). $f_d$ is then a function of the hidden state $h$ parameterized by a neural network. The probability of becoming pregnant in a cycle is the function shown at bottom.}
    \label{fig:BMS}
\end{figure}

\begin{figure}[h!]
    \centering
    \includegraphics[width=0.43\textwidth]{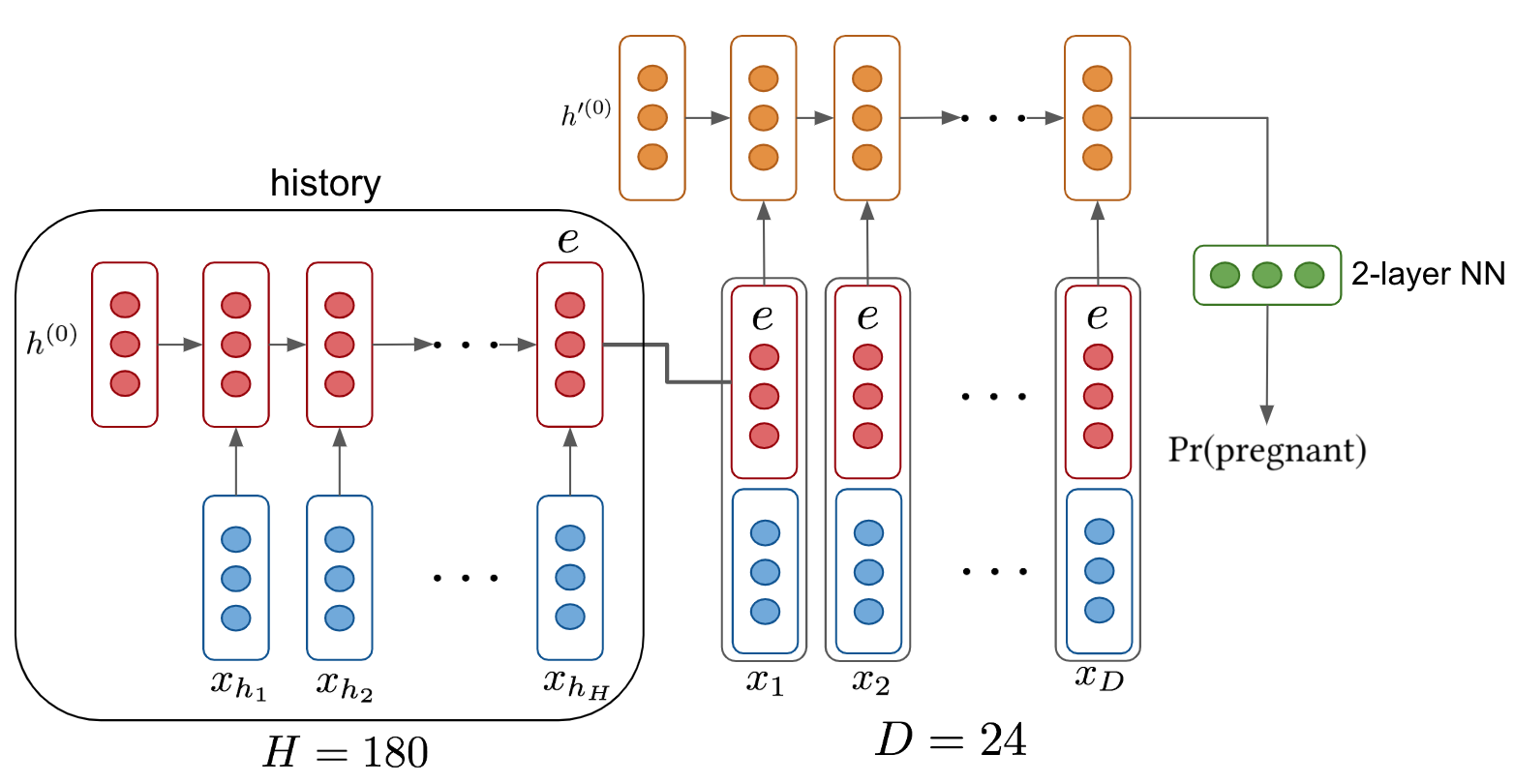}
    \caption{LSTM + user embedding: we use the history features, $x_h$, in the previous $H=180$ days for the user and feed them into the ``user history LSTM'' (shown at left). The final hidden state becomes the user embedding vector $e$, which is then concatenated with the daily features from the current cycle and fed into the pregnancy prediction LSTM (right).}
    \label{fig:embedding}
\end{figure}

\begin{enumerate} 
\item \textbf{Logistic regression}. As an interpretable baseline, we use logistic regression with coefficients for each feature and each cycle day (for example, having unprotected sex on cycle day 14), and coefficients for each user-specific feature. Our feature vector for each example consists of 2,771 features. 
\item \textbf{LSTM}. We use an LSTM network with 24 timesteps (one for each cycle day). For each time-step, we feed as input the observed features for that cycle day (binary features plus continuous features, encoded as described in Section~\ref{continuous_features_section}) along with the cycle averages for continuous features. The final hidden state of the LSTM is concatenated with the user-specific features and fed into a fully connected layer to produce the prediction. 
\item \textbf{LSTM + BMS fertility model}. We develop an LSTM-based extension of the BMS model. As described in Section \ref{fertility_modeling_lit_review}, the BMS model assumes each act of sex contributes independently to the probability a cycle results in pregnancy. Thus, we model a user's probability of becoming pregnant in a cycle as
\begin{equation*}
    \Pr(\text{pregnant}) = 1-\prod_d \prod_t \bigg(1-r_{t}\cdot f_{d}\bigg)^{s_{dt}}
\end{equation*}
where $d$ is cycle day, $t$ the type of sex (protected, unprotected, withdrawal, or none\footnote{We allow the model to learn a non-zero probability of getting pregnant even if no sex is logged to account for missing data, where users neglect to log sex but still get pregnant; this significantly improves the BMS model's predictive accuracy.}), $s_{dt}$ indicates if sex of type $t$ was logged on day $d$, $f_{d} \in (0, 1)$ is the contribution of sex on cycle day $d$ towards pregnancy, and $r_t \in (0, 1)$ is a learned parameter capturing the risk of a kind of sex (eg, $r_\text{unprotected} > r_\text{protected}$). The term inside the product is the probability that an act of sex of type $t$ on cycle day $d$ does \emph{not} result in pregnancy. We model $f_{d}$ as a function of the daily features $x_d$ using an LSTM, as illustrated in Figure~\ref{fig:BMS}. 

\item \textbf{LSTM + user embeddings}. Using only the user's current cycle does not account for their full history: eg, a user who has frequently logged unprotected sex for a long time but has not yet become pregnant may be less fertile. To incorporate information prior to the current cycle, we use a ``user history LSTM'' to encode the 180 days of user history prior to the user's current cycle, then use this LSTM's final state as a user embedding vector which we concatenate onto the other features and feed into a second LSTM as before. We jointly train both LSTMs. Figure~\ref{fig:embedding} illustrates the model architecture. 
\end{enumerate}

\begin{table*}[!h]
\small
  \caption{Predictive performance of all models. The third column provides the probability a pregnancy test is positive for users in the top 10\% vs bottom 10\% of pregnancy risk; the fourth column provides the probability over six cycles.}
  \label{predictive-performance}
  \centering
  \begin{tabular}{llp{4cm}p{4cm}}
    \toprule
    \textbf{Model}     & \textbf{AUC} & \textbf{Single cycle $p_{\text{preg}}$} &  \textbf{Six cycle $p_{\text{preg}}$} \\
      \midrule
  Logistic regression & 0.63 & 28\% vs 5\% & 86\% vs 26\%  \\ 
  LSTM & 0.65 & 30\% vs 4\% & 88\% vs 23\%  \\ 
  LSTM + BMS & 0.64 & 29\% vs 4\% & 87\% vs 22\%  \\ 
  LSTM + user embeddings & 0.67 & 30\% vs 5\% & 89\% vs 27\%  \\ 
    \bottomrule
  \end{tabular}
\end{table*}

\begin{figure*}[h]
  \centering
  \includegraphics[width=4.8in]{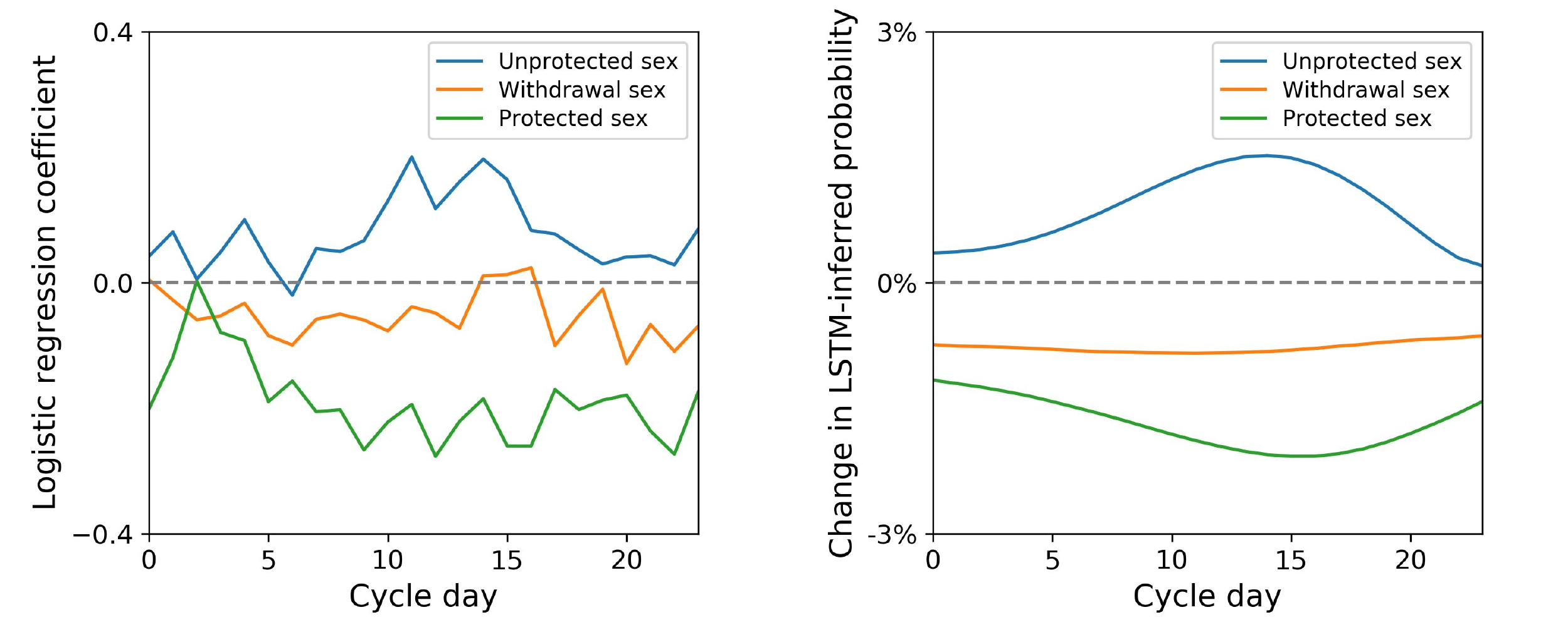}
  \caption{Model-learned time trends are interpretable for both the simple logistic regression model (left plot) and the best-performing LSTM + user embedding model (right plot). The horizontal axis is the cycle day. The vertical axis in the left plot is the logistic regression weight for logging a feature on that cycle day. The vertical axis in the right plot is how much logging a feature on a cycle day affects the LSTM-inferred probability of pregnancy. In both plots, positive y-values indicate associations with positive pregnancy tests, and negative y-values indicate associations with negative pregnancy tests. Both models learn that protected sex (green line) is negatively associated with pregnancy, while unprotected sex (blue line) is positively associated, particularly during the fertile window, and withdrawal sex (orange line) is intermediate.}
    \label{fig:interpretability}
\end{figure*}

\section{Results} 

\subsection{Predictive performance} 

\xhdr{Metrics} We evaluate model performance using AUC. Importantly, becoming pregnant is inherently a somewhat random process: it is impossible to \emph{guarantee} a woman will become pregnant in a particular cycle. Consequently, achieving a very high AUC on our task is unlikely to be feasible, and we would not expect AUC to be very high. Nonetheless, we report AUC because it is a standard metric. As a second metric, we compare the probability $p_{\text{preg}}$ that a user will log a positive pregnancy test after a cycle for the users in the top 10\% of predicted pregnancy probabilities and users in the bottom 10\% of predicted probabilities. (Note that $p_{\text{preg}}$ is the true, not the predicted, probability of a positive pregnancy test.) However, the goal of fertility counseling is not to guarantee a woman will get pregnant in a particular cycle, but rather that over some reasonable time period (eg, 6 cycles) she has a good chance of becoming pregnant if she continues her current pattern of behavior. Thus, we also compute the 6-cycle probability of becoming pregnant, assuming that each cycle contributes independently to the probability of pregnancy and that woman continues her current pattern of behavior. For example, if a woman's single-cycle $p_{\text{preg}}$ is 0.2, her six-cycle $p_{\text{preg}}$ is $1 - (1 - 0.2)^6$. 

\xhdr{Results} The LSTM with user embeddings has the highest AUC (0.67) (Table \ref{predictive-performance}). This model is able to meaningfully stratify users by their probability of getting pregnant: for example, the top 10\% of users have an 89\% chance of pregnancy over 6 cycles, whereas the bottom 10\% have only a 27\% chance. While the other three models have somewhat lower AUCs, all four models usefully stratify users by pregnancy probability. The BMS model slightly worsens LSTM performance, possibly because of the more restrictive assumptions of its probability model -- that is, that the overall probability of not getting pregnant in a cycle is the product of the probabilities of not getting pregnant on each cycle day. This demonstrates that traditional fertility models may not yield optimal predictive performance on mobile health datasets.

\subsection{Interpretability} 
\label{sec:model_coefficients}

We next assess whether our models learn \emph{interpretable} time trends which are consistent with prior fertility research. Because sexual activity is the feature most fundamentally associated with pregnancy, we examine time trends for the three types of sex logged in Clue data: unprotected, protected, and withdrawal sex. 

\xhdr{Interpreting the logistic regression model} Extracting time trends from the logistic regression model is straightforward, because the model is specifically developed to be interpretable. To understand how a feature contributes to the probability of pregnancy, we simply plot the coefficients for logging the feature on each day (Figure \ref{fig:interpretability}, left). The daily feature which is most strongly associated with positive pregnancy tests (averaging weights across all cycle days) is unprotected sex; the feature which is most strongly associated with negative pregnancy tests is protected sex. 

\xhdr{Interpreting the LSTM models} Interpreting deep learning models is much less straightforward, since they are high-dimensional and nonlinear and their coefficients do not have clear meanings. We use the following technique to quantify how logging binary feature $b$ on day $d$ influences the model's inferred probability of pregnancy: 
\begin{enumerate}
\item For each example in our test set, we set binary feature $b$ to 1 on day $d$, and compute the modeled probability of pregnancy for the example, $\Pr(\text{preg} | x_{bd} = 1)$.
\item For each example in our test set, we set binary feature $b$ to 0 on day $d$, and compute the modeled probability of pregnancy for the example, $\Pr(\text{preg} | x_{bd} = 0)$.
\item We compute $\Pr(\text{preg} | x_{bd} = 1) - \Pr(\text{preg} | x_{bd} = 0)$, averaging across examples. This corresponds to the average difference in modeled probabilities when the user does and does not log feature $b$ on day $d$. 
\end{enumerate}

Figure \ref{fig:interpretability} (right) shows results for the LSTM model with the best predictive performance (LSTM + user embeddings); results for the other two LSTM models are qualitatively similar. Like the logistic regression model, the LSTM model learns that unprotected sex (blue) is more positively associated with positive tests than protected sex (green) or withdrawal sex (orange). Unprotected sex during the middle of the cycle shows the strongest positive association. 

\xhdr{Consistency with prior research} These modeled time trends are consistent with previous fertility research, which finds that unprotected sex during the ``fertile window'' ~\citep{wilcox2000timing} is most likely to result in pregnancy. The modeled increase in pregnancy probability due to unprotected sex on a single day is fairly small (<2\% for the LSTM model); a plausible explanation for this is that many users are taking birth control. The negative weights for protected and withdrawal sex are of interest because they indicate that the associations learned by the model do not necessarily have causal interpretations. Obviously, having protected sex never reduces the chance of pregnancy in a \emph{causal} sense; rather, it indicates that the woman is trying to avoid becoming pregnant. Overall, this analysis demonstrates that we can extract interpretable, biologically plausible time trends for sexual activity, the feature most fundamentally associated with pregnancy, for both the simple logistic regression model and the predictively superior LSTM models. 

\section{Discussion} We develop four models for pregnancy prediction, combining ideas from both classical fertility modeling and modern deep learning techniques, and assess them using women's health tracking mobile data. We show we can stratify women by their probability of becoming pregnant and learn interpretable time trends consistent with prior fertility research. We develop a simple, intuitive technique to extract time trends from our LSTM models which is more broadly applicable to other time series datasets. 

Based on our study, we recommend two steps to enable women's health tracking apps to reach their full potential for pregnancy prediction. First, companies should fully describe the algorithms used for pregnancy prediction; this will facilitate constructive criticism and increase confidence in the algorithms. Second, incentivize users to provide more complete data. In the Clue dataset, most users do not log pregnancy tests or predictively useful features like BBT; encouraging users to provide this data will improve predictions. 

These two improvements can facilitate prediction tasks which build on the one described here. For example, early identification of people who are very unlikely to become pregnant can allow faster referrals to infertility treatment. Modeling the behaviors which \emph{causally} contribute to pregnancy can be used to recommend behavior: for example, having sex during a person-specific fertile window. Progress on these prediction tasks will improve well-being for the millions of people pursuing pregnancy. 

\xhdr{Acknowledgments} We thank Paula Hillard, Pang Wei Koh, Chris Olah, Nat Roth, and Camilo Ruiz for helpful comments; Amanda Shea at Clue for data assistance; all Clue users who volunteered their data for research; and the Hertz, NDSEG, and Swiss National Science Foundations for funding. This research was supported in part by the Stanford Data Science Initiative and the Chan Zuckerberg Biohub.

\clearpage

\bibliographystyle{ACM-Reference-Format}
\balance
\bibliography{main}

\end{document}